\documentclass{aa}  

\usepackage{graphicx}
\usepackage{color}
\usepackage{txfonts}
\usepackage{amsmath}
\usepackage[]{hyperref}
\begin{document}

   \title{Mutual neutralization of C$_{60}^+$ and C$_{60}^-$ ions}
   \titlerunning{Mutual neutralization of C$_{60}^+$ and C$_{60}^-$ ions}
   \subtitle{Excitation energies and state-selective rate coefficients}

   \author{
   Michael Gatchell\inst{1} \and
   Raka Paul\inst{1} \and
   MingChao Ji\inst{1} \and
   Stefan Ros\'en\inst{1} \and 
   Richard D.\ Thomas\inst{1} \and
   Henrik Cederquist\inst{1} \and
   Henning T.\ Schmidt\inst{1} \and
   {\AA}sa Larson\inst{1} \and
   Henning Zettergren\inst{1}
  }

   \institute{Department of Physics, Stockholm University, 106 91 Stockholm, Sweden
              \email{gatchell@fysik.su.se} 
                      }

   \date{Received September 15, 1996; accepted March 16, 1997}

 
  \abstract
   {Mutual neutralization (MN) between cations and anions plays an important role in determining the charge balance in certain astrophysical environments. However, empirical data for such reactions involving complex molecular species have been lacking due to challenges in performing experimental studies, leaving the astronomical community to rely on decades-old models with large uncertainties for describing these processes in the interstellar medium.}
   {Our aim is to investigate the MN reaction C$_{60}^+$ + C$_{60}^-$ $\rightarrow$  C$_{60}^*$ + C$_{60}$ for collisions at interstellar-like conditions.}
   {We studied the MN reaction between C$_{60}^+$ and C$_{60}^-$ at collision energies of 100\,meV using the Double ElectroStatic Ion Ring ExpEriment (DESIREE) and its merged beam capabilities. To aid in the interpretation of the experimental results, semiclassical modeling based on the Landau-Zener approach was performed for the studied reaction.}
   {We experimentally identified a narrow range of kinetic energies for the neutral reaction products. Modeling was used to calculate the quantum state-selective reaction probabilities, absolute cross sections, and rate coefficients of these MN reactions, using the experimental results as a benchmark. We compared the MN cross sections with model results for electron attachment to C$_{60}$ and electron recombination with C$_{60}^+$.}
   {Our results show that it is crucial to take mutual polarization effects, the finite sizes, and the final quantum states of both molecular ions into account in order to obtain reliable predictions of MN rates expected to strongly influence the charge balance and chemistry in environments such as dense molecular clouds.
}


   \maketitle
%

\section{Introduction} \label{sec:intro}
To date, more than three hundred molecular species have been identified in space through their unique spectroscopic fingerprints \citep{McGuire:2022aa,Muller:2001aa}. These include polycyclic aromatic hydrocarbons (PAHs) and fullerenes (C$_{60}$ and C$_{70}$), which are the hitherto largest molecular types. PAHs are believed to be key players in star and planet formation and in the evolution of galaxies, acting as seeds for carbonaceous grains and nucleation sites for chemical reactions \citep{Tielens2013}. The first unambiguous identifications of PAH-like species in space was recently reported from radio observations of the Taurus molecular cloud \citep{McGuire2021,Cernicharo2021}. Since the discovery of neutral C$_{60}$ molecules in a young planetary nebula \citep{Cami:2010aa} and in reflection nebulae \citep{Sellgren_2010}, these molecules have been detected in a wide a range of astrophysical environments \citep{Berne2017}. In 2015,  C$_{60}^+$ ions were shown to be responsible for a few of the several hundreds of so-called diffuse interstellar bands (DIBs) \citep{Campbell:2015aa}. After a century of speculation and attempts to understand the origin of DIBs, this is still the only firm identification of a specific DIB carrier \citep{Campbell:2015aa,Linnartz:2020aa}. More recently, C$_{60}^-$ was tentatively identified, together with C$_{60}$ and C$_{60}^+$, in the star-forming region of the Perseus molecular cloud, indicating relative abundances of 1:7:2 for C$_{60}^-$:C$_{60}$:C$_{60}^+$ \citep{Iglesias-Groth:2019}.

The charge balances observed in local astrophysical environments reflect their various thermal and nonthermal reaction processes. These balances change dramatically when moving from, for example, the intense radiation fields near hot central stars through the photo dissociation regions and into the lower temperature environments of molecular clouds where stars are born. As PAHs and fullerenes are most likely omnipresent in space in neutral and various charged forms, their charge balances may be used to calibrate astrophysical models of the evolution of molecules and their roles in such processes as star and planet formation. However, uncertainties in model input data and missing key reactions, such as mutual neutralization (MN), may significantly influence the understanding of the chemical and physical properties of studied regions.
For instance, molecular anions are predicted to be abundant in photo dissociation regions according to astrophysical models \citep{Millar_2017} but have, to the best of our knowledge, not yet been observed in such environments. A possible reason for this is that MN is much more efficient at reducing the anion population than assumed in models, which typically use a common single-valued rate coefficient for all MN reactions in a given environment (see \citet{Millar_2017} and references therein). Another example is dense molecular clouds where PAH anions are assumed to replace free electrons as the dominant negative charge carriers \citep{Lepp_1988,Wakelam_2008}. MN reactions between PAH anions and atomic cations could then play a leading role in reducing the overall ionization fraction in such clouds \citep{Wakelam_2008}. So far, calculations of molecular MN rate coefficients have often been based on a model developed in the 1940s to describe ion interactions with spherical dust grains \citep{Spitzer1941}. In this model, only the pure Coulomb part of the attraction between the ions is taken into account, and the electron transfer distances are taken to be similar to the geometrical sizes of the grains \citep{Wakelam_2008}. It has recently been established that ab initio and semiclassical descriptions of MN of atomic anions and cations can be used to predict the total MN rates and final state population distribution for a broad range of collision energies extending down to the sub-electronvolt regime \citep{dickinson1999,Launoy_2019, Eklund2021,Grumer2022,Poline2022,Dochain2023, SchmidtMay2023,hornquist2022}. Models describing interactions involving molecular species are, however, much less developed.

Studying astrophysical MN processes with molecules in the laboratory is challenging since hot ions from an ion source, which can have high internal energies (tens of electronvolts), have to be pre-cooled or cooled in ion-beam storage devices before one may proceed with measurements of relevance for cold interstellar environments. The Double ElectroStatic Ion Ring ExpEriment (DESIREE) ion-beam storage ring facility \citep{thomas2011double,Schmidt:2013aa} is designed for studying molecular MN reactions at unprecedented detail under conditions mimicking those in, for example, the interstellar medium. The long time storage capabilities of two merged ion beams in a cryogenic environment at an extremely low residual gas density allow for fine control of the ion-ion collision energy and for ro-vibrational cooling of molecular ions \citep{Schmidt_2017, Poline_2024,bogot2024}.

In this paper, we present experimental results on the final state distribution in C$_{60}^*$ for C$_{60}^+$ + C$_{60}^-$ $\rightarrow$  C$_{60}^*$ + C$_{60}$ MN reactions at a center-of-mass collision energy of $100\pm 50$\,meV. This is, to the best of our knowledge, the first ever experimental study of an MN reaction involving two oppositely charged, complex molecular ions under such well-defined conditions. The fullerenes serve as key model systems, as their mutual interactions can be well approximated by one-dimensional potential energy curves that simplify the model calculations, and thus, direct comparisons with the experiments can be made. Furthermore, the high stability of C$_{60}$ molecules ensures that the excitation energy following electron transfer is not sufficient to break their cages, nor is the center-of-mass collision energy gained through the C$_{60}^+$ + C$_{60}^-$ attraction sufficient to form or break bonds, even in head-on collisions \citep{Jakowski2010}. This makes the interpretation of the results much more straightforward than what is expected for MN processes involving other more reactive and/or less stable complex molecular systems. Fullerenes are omnipresent in a range of astronomical environments and are more likely to form anions than PAHs of similar sizes ($\sim$7\,a$_0$ in diameter) given the high electron affinities of fullerenes \citep{Huang2014}. Through direct measurements and modeling of the excited-state distribution of C$_{60}^*$,  following C$_{60}^+$ + C$_{60}^-$ $\rightarrow$ C$_{60}^*$ + C$_{60}$ reactions at sub-electronvolt collision energies, we arrive at semi-empirical expressions for total and state-selective MN rate coefficients as functions of temperature, which can be applied directly in astrophysical modeling work.

\begin{figure*}[]
\center
\includegraphics[width=.9\textwidth]{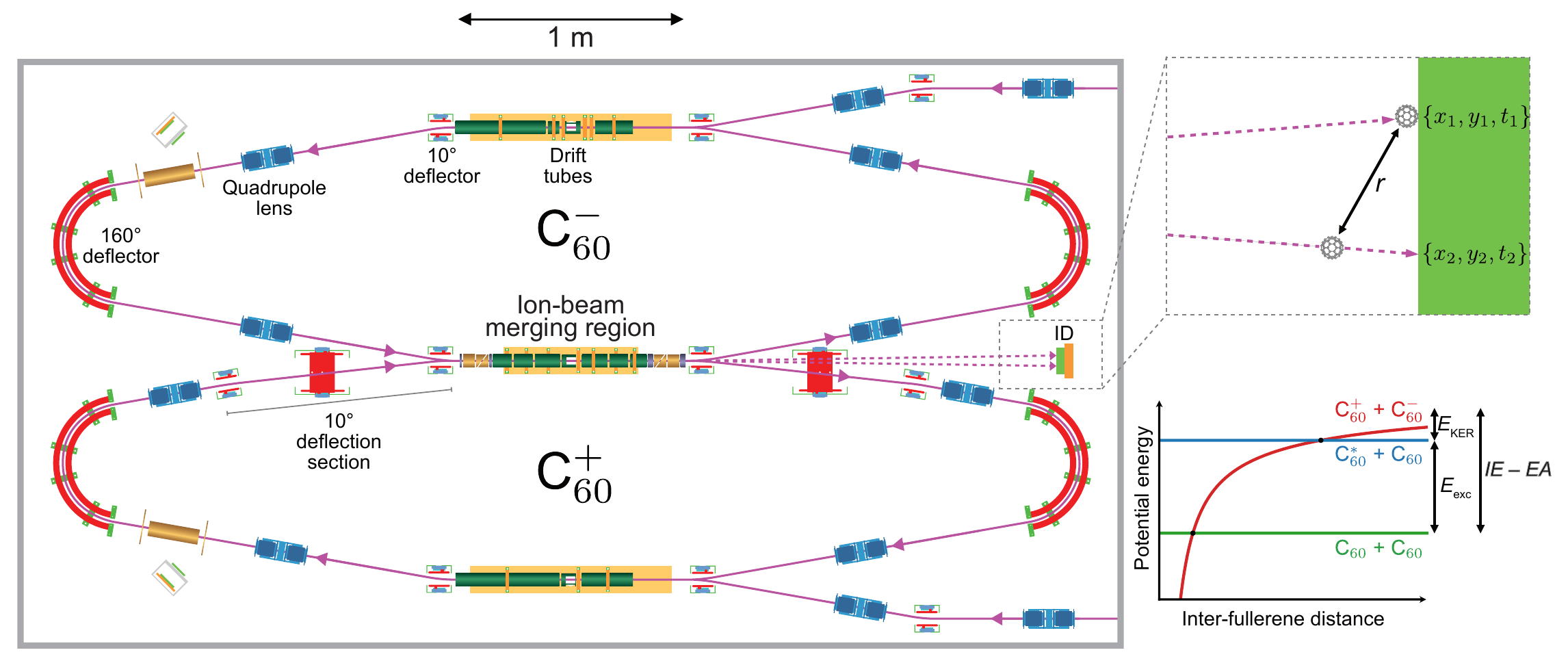} 
 \caption{Schematics of the DESIREE storage rings and the method to determine the final states in MN reactions. The left panel shows an overview of the two storage rings and the common merging section, where MN reactions may be studied with precise control of the center-of-mass collision energy. The imaging detector, marked as "ID," provides information on the separation ($r$) of the two neutrals formed in MN reactions, as shown in the upper-right zoom-in. The $r$ value depends on the kinetic energy release ($E_{\mathit{KER}}$) in the MN reaction, which is related to the excitation energy ($E_{\mathit{exc}}$) of the neutral products through the potential energy curves for the initial and final states shown schematically in the lower-right panel.}
  \label{fig:fig1}
\end{figure*}

\section{Methods}

\subsection{Experiments}
The experiments were carried out at the DESIREE ion-beam storage ring facility. We briefly summarize its main features here, as they have been described in detail elsewhere \citep{thomas2011double,Schmidt:2013aa}. A schematic of the setup is shown in Fig.\ \ref{fig:fig1}. It consists of two electrostatic ion-beam storage rings with a common straight section where oppositely charged ion beams with kiloelectronvolt energies are made to interact in a region equipped with a set of drift tubes. These allow for deceleration of one beam and acceleration of the other such that the relative velocity of the ion pairs are mainly limited by small beam divergences and any small angle between the two beams as they interact. Thereby, it is possible to study MN reactions, A$^+$ + B$^-$ $\rightarrow$ A + B, with fine control of the center-of-mass collision energy down to about 100\,meV. In the present experiments, two of the seven drift tubes were biased at 503\,V to allow the circulating 11\,keV C$_{60}^+$ and 10\,keV C$_{60}^-$ beams to interact along a 16\,cm long region at a center-of-mass collision energy of $100\pm 50$\,meV, as determined by optimizing the overlap of the two ion beams using the procedure established in previous experimental studies with atomic ions \citep{Eklund2021,Grumer2022}. This collision energy corresponds to a kinetic temperature of 10$^3$\,K, which is comparable to that found in, for example, dense photo dissociation regions \citep{RevModPhys.71.173}. The MN reactions taking place in this biased section of the drift tubes are distinguished from reactions occurring in non-biased regions by the significantly larger differences in arrival times for pairs of neutral products formed in the non-biased segments of the merging region due to the higher interaction energies there. DESIREE is operated at 13\,K and with a residual gas density of about 10$^{4}$\,cm$^{-3}$, corresponding to a background pressure of $\sim$ 10$^{-14}$\,mbar with H$_{2}$ molecules as the main component of the residual gas. This allows for the storage of ions for hours \citep{Backstrom2015}.

In the present experiments, C$_{60}$ powder (98\% purity from Sigma Aldrich) was heated in two separate ovens in order to bring the molecules into a gas phase and for them to flow into the ion sources used in the experiments. The fullerene cations were produced with an electron cyclotron resonance (ECR) ion source, while the anions were produced using an electron attachment source based on the design by \citet{Yamada2018}. The ions from the two sources were accelerated, and the mass-to-charge was selected by bending magnets along two separate injection beam-lines. The ion-beams were chopped into pulses of lengths set to fit the circumferences of the rings (8.7\,m) and then stored in the two rings. Neutrals formed in the merging section from MN processes, spontaneous decay of hot ions from the ion source, or collisions with rest gas were monitored as a function of storage time by the so-called imaging detector assembly consisting of a triple-stack microchannel plate with a phosphor screen anode from which the emitted light is detected outside the rings by means of a 16 channel photomultiplier tube and a CMOS camera \citep{Rosen2007}. This allowed the positions and time differences of the two neutrals to be determined as they hit the detector and hence the separation between them (see Fig.\ \ref{fig:fig1}). The separation, $r$, is related to the kinetic energy release, $E_{\mathit{KER}}$, through the following relation  \citep{Eklund2021}:    
\begin{equation}
r= \frac{L}{v}\sqrt{2\frac{E_\mathit{K}}{\mu}},
\end{equation}
where $E_K$ is the total kinetic energy of the two products in the center-of-mass system and is given by the sum of the collision energy in the center-of-mass system ($E_{\mathit{CM}}$) and the kinetic energy release ($E_{\mathit{KER}}$), including any ro-vibrational excitations in the MN process. The velocity of the mass center of the collision system in the laboratory system of reference is $v$, while $\mu$ is the reduced mass, and $L$ is the distance from the position where the MN reaction took place to the detector and is taken to be the mean distance from the biased drift tube region to the detector. For a given velocity of the center of mass of the particles in the laboratory system of reference, $r$ depends on the collision energy in the center-of-mass system ($E_{\mathit{CM}}$) and the kinetic energy release ($E_{\mathit{KER}}$) in the MN reaction such that $r \propto \sqrt{E_K}$, where  $E_K=E_{\mathit{CM}}+E_{\mathit{KER}}$. In turn, $E_{\mathit{KER}}$ is related to the difference in internal energies of the parent and product species. Thus by measuring the $r$ distances, the kinetic energy released in the reactions can be determined, which in turn provides information on the final states populated in the MN reactions (see Fig.\ \ref{fig:fig1}).

The energetics of the MN process are illustrated in the lower-right panel of Fig.\ \ref{fig:fig1}, which shows a schematic of the potential energy curve for the incoming ion pair (C$_{60}^+$ + C$_{60}^-$) as a red line and two examples of hypothetical combinations of final states of the two neutral C$_{60}$ molecules after the collision. The lower (green) line illustrates a situation where both neutralized molecules are in their electronic ground states, while one of them is electronically excited (C$_{60}^*$) in the upper (blue) state. In this schematic picture, the ionic precursors are in their electronic ground states, and there is no change in ro-vibrational excitation energy upon electron transfer. For the case where $E_{\mathit{CM}}$ = 0\,eV, the electronic excitation energy is $E_{\mathit{exc}}=(\mathit{IE}-\mathit{EA})-E_{\mathit{KER}}$, where $\mathit{IE} = 7.6$\,eV \citep{Muigg_1996} and $\mathit{EA} = 2.68$\,eV \citep{Huang2014} are the ionization energy and electron affinity of C$_{60}$, respectively. Thus, the measurements yield the final distribution of electronically excited states of the MN reaction products and their respective branching fractions. Such experiments have been successfully carried out for a range of atomic MN reactions in DESIREE, where individual final states are typically well resolved \citep{Eklund2021,Grumer2022,Poline2022,SchmidtMay2023}. 

It is essential to be able to store the ions for times longer than about 100 milliseconds, as the spontaneous unimolecular decay of internally hot C$_{60}^-$ ions at earlier times give rise to neutral products \citep{Gatchell:2024aa}, which makes it harder to identify MN processes. However, at timescales longer than this, the population of ions has cooled enough to not decay spontaneously. This together with the excellent vacuum conditions in DESIREE allow the MN reactions with (complex) molecular systems to be studied in unprecedented detail.

\subsection{Model calculations}

The present experimental results were used to benchmark semiclassical multichannel Landau-Zener model calculations. The model posits that electron transfer between oppositely charged fullerene ions leads to electronic excitation of the electron-accepting fullerene, while the electron-donating fullerene is left in its electronic ground state. Vibrational excitations are neglected, given the similar structural configurations of the neutral and ionic fullerene molecules.

Provided that the electronic coupling between the states associated with the ion pair and neutral fragments can be correctly estimated, the model has in many cases been successfully used for studies of MN in collisions between atomic ions (see, e.g.,\ refs.\ \citep{Bates_1955,Hedberg_2014, Barklem_2021,Liu_2023}). Applications of the Landau-Zener model to collisions of molecular ions are rare. Examples are studies of MN in collisions of H$^-$ with H$_2^+$ \citep{Eerden_1995} and H$_3^+$ \citep{Janev_2006} as well as Ar$^+$ with SF$_n^-$, $n=6,5,4$~\citep{Bopp_2008}.

The Landau-Zener model~\citep{landau_1932,zener_1932} is a two-state model where the probability, $P$, for a diabatic passage through a curve crossing is given by 
\begin{equation}
P= e^{-w},
\end{equation}
where
\begin{equation}
w= \frac{2 \pi H^2_{12}}{v_r|\frac{d} {dR}(U_i-U_{vdW})|}.
\end{equation}
Here, $H_{12}$ is the electronic coupling element between the ionic and van der Waals state, $v_r$ is the radial velocity, and $\frac{d} {dR}(U_i-U_{vdW})$ is the difference of slopes of the ionic and van der Waals state potentials. All properties are evaluated at the curve crossing distance $R_c$.

In this study, we use the so-called Girifalco potential \citep{Girifalco} for the van der Waals states ($U_{\mathit{vdW}}$) between two neutral fullerene molecules. This is a Lennard-Jones type potential, where the sixty carbon atoms of a fullerene are approximated as being spread out on the surface of a sphere and interacting with another such model system. This potential has been shown to reproduce inherent properties of bulk fullerite materials as well as clusters of fullerenes remarkably well \citep{Hansen:2022}. This includes the binding distance (18.9\,a$_0$) and the binding energy (0.28\,eV) of the van der Waals fullerene dimer, with the latter in excellent agreement with experimental results \citep{Branz2002}. As these parameters do not change significantly when the dimer system is ionized, we assumed the same applies for the excited van der Waals states. We thus simply shifted the Girifalco potential by the excitation energy of C$_{60}^*$ for the C$_{60}$ + C$_{60}^*$ system at infinite separation in order to define the potential energy curves for all accessible van der Waals states. Here, we have calculated the ground state and excited (singlet and triplet) states using time-dependent density functional theory (TD-DFT) calculations using the PBE0 functional~\citep{PBE0} and the 6-311++G(d,p) basis set. The calculated vertical excitation energies are provided in the Appendix. In the Landau-Zener calculations, we shifted the calculated energy values for the excited states by +0.15\,eV to better reproduce results from experiments and more advanced calculations using the ab initio cluster expansion methods \citep{Leach:1992aa,Catalan:1994aa,Smith:1996aa,Fukuda:2012aa} and hence correct for systematic uncertainties associated with TD-DFT calculations.

To determine the ion-pair potential ($U_i$) of the C$_{60}^+$ + C$_{60}^-$ interaction, we used a superposition of the attractive part of the Girifalco potential and the interaction potential of two oppositely charged rigid metal spheres. Fullerenes perturbed by external charges have been shown to be accurately described by such a simple electrostatic model, as demonstrated through comparisons with DFT calculations of ionization energy sequences and the interaction energy of a fullerene with a point charge \citep{Zettergren2012}. The model has also been successfully used to reproduce experimental results on the stabilities of (multiply) charged fullerene dimers and larger clusters \citep{Zettergren2007} as well as charge transfer reactions in collisions between multiply charged atomic or fullerene ions with neutral fullerenes at kiloelectronvolt energies \citep{Zettergren2002}. This is due to the many delocalized electrons and their high mobility on the spherical surface such that it behaves as an infinitely conducting macroscopic metal sphere when it is polarized. An analytical expression for the interaction energy of two oppositely charged metal spheres can then be derived using the methods of electrostatic image charges, where the mutual polarization is given by an infinite number of image charges in each sphere whose magnitude decreases and converges to a specific value for a given inter-fullerene separation $R$. We used 100 image charges, as including more has no effect on the potential energy curves for the shortest $R$ distances that are relevant for the LZ modeling in the present work. The only free parameter in this model is the fullerene radius, which we set to $a=8.6$\,a$_0$ based on comparisons with DFT results \citep{Zettergren2012}. This radius is larger than the cage radius (6.7\,a$_0$), as it includes the spill-out of the electron cloud and is in reasonable agreement with the experimental polarizability value ($a^3=(76.5 \pm 8)$\,\AA$^3$ \citep{Ballard2000}), yielding $a = (8.0 \pm 0.3)$\,a$_0$ \citep{Antoine:1999aa}.

We estimated the electronic couplings between the ionic and van der Waals states using the semi-empirical formulas developed by \citet{olson_1972} to accurately describe one-electron transfer processes. The coupling elements are given by
\begin{equation}
    H_{ij}=\sqrt{E_i\cdot E_f}1.044R^{*}\exp(-0.857R^{*}),
\end{equation}
where $E_i=\mathit{EA}$ is the electron affinity of the transferred electron and $E_f=\mathit{IE}-E_{\mathit{exc}}$ is the effective ionization energy of the transferred electron in the final C$_{60}^*$ state. The scaled crossing distance $R^*=\frac{1}{2}\left(\alpha+\gamma\right)R_c$ is introduced with $\alpha=(2E_i)^{1/2}$ and $\gamma=(2E_f)^{1/2}$. All quantities are in atomic units. 

The branching ratios and total MN cross section are calculated using the multichannel Landau-Zener model, where the probabilities for neutralization are obtained by adding the probabilities from all possible pathways as described by \citet{Salop_1976}. Calculations are done separately on the singlet and triplet manifolds of states, and the statistical weights are considered (one-fourth and three-fourths for the singlet and triplets states, respectively) when calculating the branching ratios.  It should be noted that there is a large density of van der Waals states with close lying potential crossings with the ion-pair potential. Still, due to the large reduced mass of the fullerene dimer, the Landau-Zener transition regions estimated as $v_r\tau_{\mathit{jump}}$, where $v_r$ is the radial velocity at the curve crossing and $\tau_{\mathit{jump}}$ is the Landau-Zener transition time~\citep{vitanov_1999}, are smaller than the distances between neighboring curve crossings. This justifies the use of the multichannel Landau-Zener model.

The calculated MN cross section was parameterized in terms of the initial relative velocity $v$ between the fullerene ions at infinite separation as 
\begin{equation}\label{eq.x}
    \sigma_{\mathit{MN}}(v)=Av^{-2}+Bv^{-1}+C+Dv.
\end{equation}
A least-squares fit to the calculated MN cross sections in the energy range 0.001--100\,eV resulted in the fitted parameters given in the Appendix.
An expression for the thermal MN rate coefficient was obtained by integrating the fitted cross section over a Maxwellian velocity distribution (eq.\ \eqref{eq:rate}).

\subsection{Simulations of excitation energy distributions}
We have developed a method to simulate and analyze the measured $r$ distributions from MN experiments at DESIREE where the experimental parameters are taken into account, such as the kinetic energy spread of the two stored beams, the beams’ divergences, the merging angle of the two beams, and the electric potential around the biased drift tubes. A detailed description of this Monte Carlo approach can be found in \citet{Eklund:2020aa}. In the present simulations, the electric potential in the drift tube region was taken from SIMION 8.1 simulations, while the other parameters listed above were calibrated during earlier successful MN experiments when only atomic collision partners were involved \citep{Eklund2021,Grumer2022}. 

For a given value of $E_{\mathit{KER}}$, a number of collision events were randomly generated upon the experimental parameters in the merging section, and then the center-of-mass collision energy $E_{\mathit{CM}}$ of each ion pair was calculated. The sum $E_{K} = E_{\mathit{CM}} + E_{\mathit{KER}}$ is shared by the two neutral products after the collision. In the studied energy range of $E_{\mathit{CM}} = 100\pm50$\,meV, the angular distribution of the two neutrals is approximately isotropic, and with the help of a random direction vector, the total velocity vectors of the neutral products could be obtained. The arrival time difference between the two neutrals and their distance on the imaging detector, labeled "ID" in Fig.\ \ref{fig:fig1}, could then be calculated in the laboratory frame. The separation $r$ of the two neutrals as they passed the detector plane was calculated in the same way as for the experimental data. By repeating this procedure for a large number of randomly generated experimental parameters for the different possible $E_{\mathit{KER}}$ values, corresponding $r$ distributions were obtained.

\begin{figure}[]
\center
\includegraphics[width=1\columnwidth]{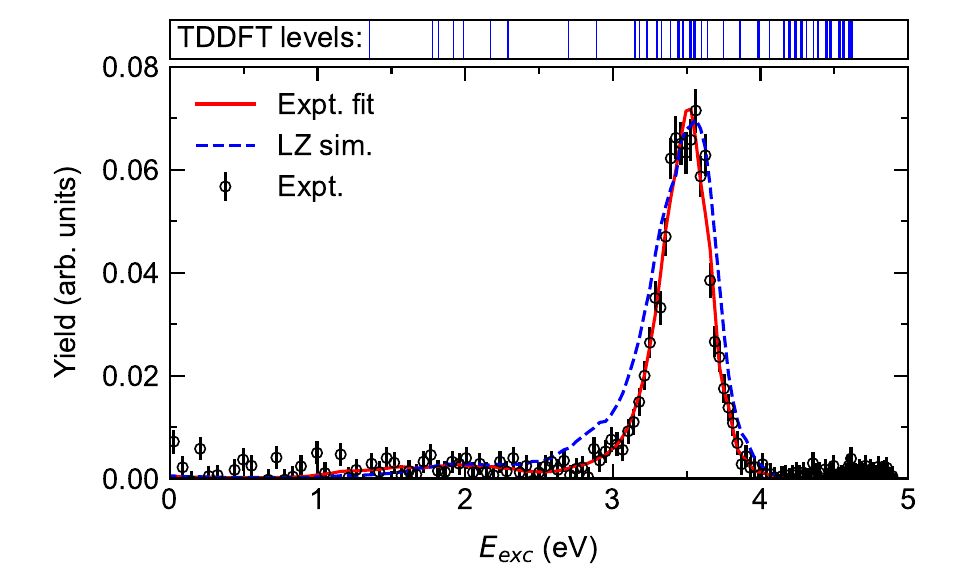} 
 \caption{Measured and simulated distributions of the final excitation energies of the two neutral reaction products formed in $100$\,meV C$_{60}^+$ + C$_{60}^-$ $\rightarrow$  C$_{60}^*$ + C$_{60}$ collisions. The measured $E_{\mathit{exc}}$ distribution (black circles) reflects the kinetic energy release in the MN process. The blue dashed curve is the simulated result based on input from the present Landau-Zener model calculations when mutual polarization effects and the finite sizes of the collision partners are taken into account. The blue lines in the upper panel display the available final energy levels according to our TD-DFT calculations.}
  \label{fig:fig2}
\end{figure}

\begin{figure*}[]
\center
\includegraphics[width=1\textwidth]{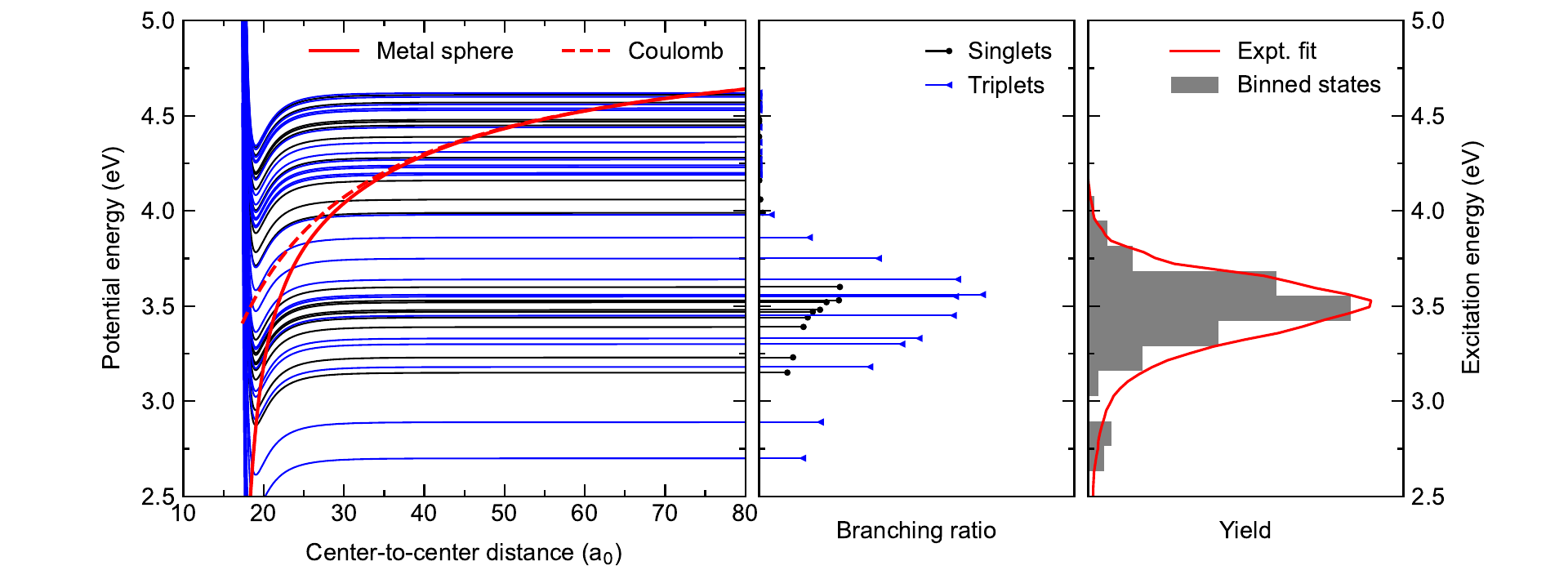} 
 \caption{Model potential curves and final state populations. The left panel shows the diabatic potential energy curves used in the present Landau-Zener model calculations. The solid red curve is for the ion-pair state and is a superposition of the interaction energy of two oppositely charged rigid metal spheres and long-range van der Waals interaction, while the dashed red curve shows the interaction energy for oppositely charged point charges. The potentials of the van der Waals states in black (singlet states) and blue (triplet states) are described using the Girifalco potential. These are separated by the excitation energies of C$_{60}$ at infinite separation, as calculated by means of TD-DFT calculations. The middle panel shows the predicted relative final-state populations from the multichannel Landau-Zener model calculation, which predicts that a broad band of states are populated and is in agreement with the experimental findings. In the right panel, the fitted yield of excitation energies obtained from the experimental fit (red curve) is compared with the binned branching ratios obtained from the modeling.}
  \label{fig:fig3}
\end{figure*}

\section{Results}
For complex molecular systems such as fullerenes, there are typically many close-lying bands of states in the final products, as indicated by the calculated energy levels shown at the top of Fig.\ \ref{fig:fig2}. Depending on which states get populated in the MN process, this gives a broad final $E_{\mathit{exc}}$ distribution, as seen in the lower panel of Fig.\ \ref{fig:fig2}, where the black circles are the present experimental results. The width of the measured distribution corresponds to a spread in the excitation energy in the reaction products of about 0.5\,eV full width half maximum. We did not observe any significant changes in the peak position or the peak width in the present range of storage times, which span from 100 milliseconds to 10 seconds, nor in the total MN yield. This suggests that the same band of final states are populated during the entire measured time range, which is consistent with efficient radiative cooling of vibrationally hot fullerene cations and anions occurring on timescales below 100 milliseconds through the process of recurrent fluorescence, as reported in earlier studies \citep{Hansen1996,Andersen1996}. On longer timescales, the molecules may continue to cool slowly via IR emission due to transitions between vibrational states, but this slow cooling has no appreciable effect on the measured distribution of $E_{\mathit{exc}}$.

In Fig.\ \ref{fig:fig3}, we show the model potential energy curve for the ion-pair state where the long range interaction is dominated by that of two oppositely charged rigid metal spheres (red solid line). This is compared to the dashed curve showing the interaction energy between two point charges. At long ranges, the two curves converge, but at distances below a few tens of Bohr radii, the additional effects included in the metal sphere model dominate the interaction. It is at these shorter distances that the majority of MN reactions take place according to the present measurements and calculations. The black and blue curves are the potential energy curves for all accessible van der Waals singlet and triplet states, respectively, according to our model calculations (see the Methods section for details). In the middle panel of Fig.\ \ref{fig:fig3}, the calculated Landau-Zener reaction probabilities (branching ratios) are shown for when we use the metal sphere model for the ion-pair state and a center-of-mass collision energy of $E_{\mathit{CM}} = 100$\,meV. The modeling suggests that a broad band of van der Waals states are populated, with excitation energies centered around $E_{exc}=3.6$ eV ($E_{\mathit{KER}}=1.3$ eV). 
In the right panel of Fig.\ \ref{fig:fig3}, we have binned the combined branching ratios of singlet and triplet states into the gray bars. As seen in this panel, this distribution compares well with the present experimental results (the red curve in Fig.\ \ref{fig:fig3} is identical to the red-curve fit to the experimental data in Fig.\ \ref{fig:fig2}).

 The resulting model distribution, convoluted with the simulated experimental response function, is shown as a blue dashed line in Fig.\ \ref{fig:fig2}. The peak position is in close agreement with the experimental results, although the calculated distribution has a slightly larger width. 
We note that the potential energy curves of the states populated in the MN process cross the ion-pair state at distances where it is crucial to consider the mutual polarization and the finite sizes of the collision partners. This shows that it is important to take polarization and size effects, as well as the features of the excitation energy spectrum of the neutral products, into account when modeling MN reactions involving complex molecules such as fullerenes.

\begin{figure}[]
\center
\includegraphics[width=1\columnwidth]{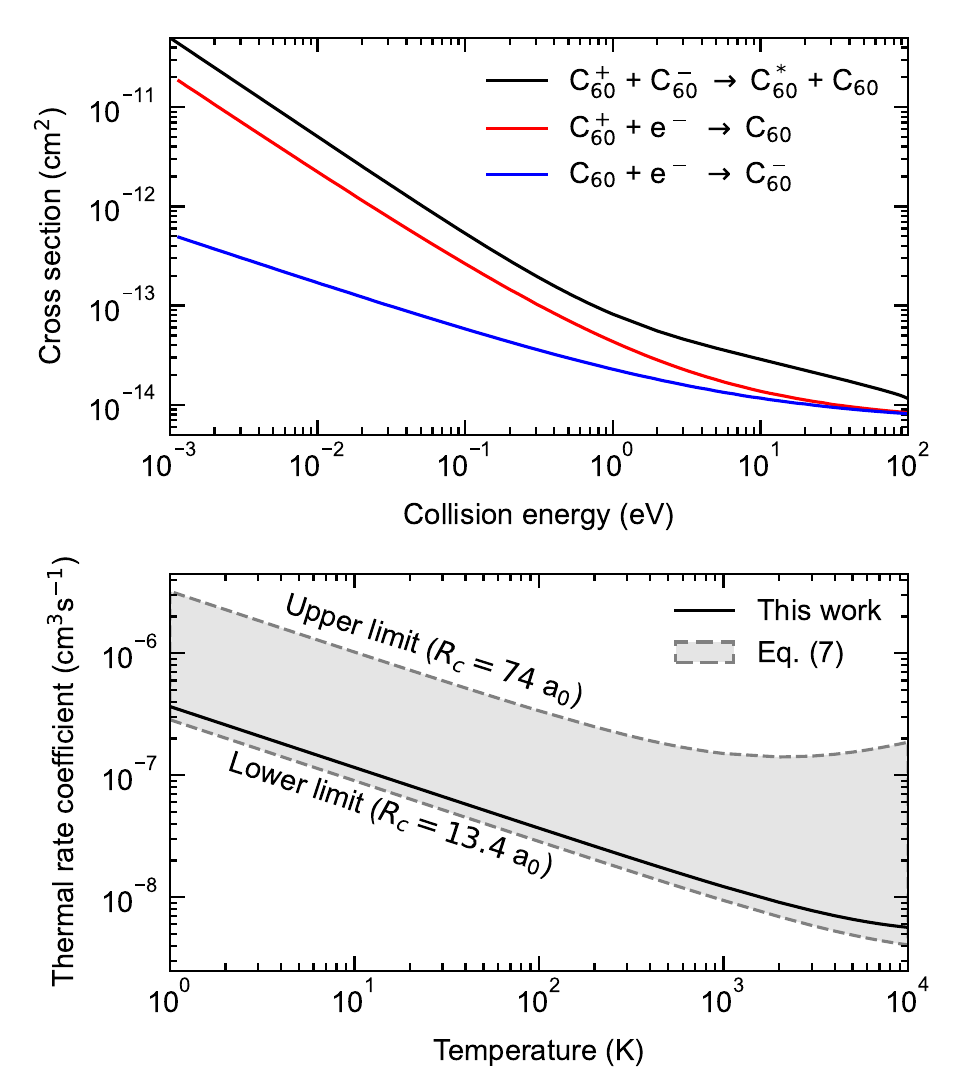} 
 \caption{Semi-empirical MN cross section and thermal rate coefficient as functions of center-of-mass collision energy and temperature, respectively. The top panel shows calculated total cross sections for C$_{60}^+$ + C$_{60}^-$ $\rightarrow$  C$_{60}^*$ + C$_{60}$ (black), C$_{60}^+$ + e$^-$ $\rightarrow$ C$_{60}$ (red), and C$_{60}$ + e$^-$ $\rightarrow$ C$_{60}^-$ (blue) reactions. The MN thermal rate coefficient is shown in the bottom panel with the black curve. There, the gray area shows the ranges of MN rates obtained using a simple model developed for the neutralization of dust particles in the interstellar medium \citep{Spitzer1941}, which has previously been used to estimate MN rates involving PAH anions in dense molecular clouds \citep{Wakelam_2008}.}
  \label{fig:fig4}
\end{figure}

As the simple metal sphere model reproduces the measured final state distribution when these effects are taken into account, and hence captures the essence of the present MN reaction, we used this model to calculate the absolute MN cross section as a function of collision energy. The result is shown with the black curve in the top panel of Fig.\ \ref{fig:fig4} together with the cross sections for electron recombination, C$_{60}^+$ + e$^-$ $\rightarrow$ C$_{60}$ , and electron attachment, C$_{60}$ + e$^-$ $\rightarrow$ C$_{60}^-$, which were calculated using classical capture theory following \citet{Linden2016}. The cross section for the MN reaction is larger than for electron recombination, especially at low but also at high collision energies, where the two cross sections approach their respective geometrical cross sections. The electron attachment cross section is significantly lower than those for MN and electron recombination at low collision energies, but it converges to that of electron recombination at collision energies above about 100\,eV.  

The thermal rate coefficients for C$_{60}^+$ + C$_{60}^-$ MN, electron recombination with  C$_{60}^+$, and electron attachment to neutral  C$_{60}$, were obtained by integrating the cross sections over an isotropic Maxwellian velocity distribution. The thermal rate coefficients are given by 
\begin{equation}
\begin{aligned}
\label{eq:rate}
k(T) &= \int_0^{\infty} \sigma v f(v,T) 4\pi v^2 dv= \\ &= \alpha_0+\alpha_1(T/300)^{1/2}+\alpha_2(300/T)^{1/2}+\alpha_3(T/300),
\end{aligned}
\end{equation}
 where the temperature, $T$, is in Kelvin and $v$ is the relative velocity of the reactants. The fit parameters $\alpha_i$ ($i=0-3$) are given in the Appendix. The MN rate is shown as a black solid line in the right panel of Fig. 4, and it follows a $T^{-1/2}$ behavior up to about 1000\,K. As illustrated below, this is typical for an interaction potential that is dominated by a long-range Coulomb attraction \citep{Spitzer1941}.

\section{Discussion and conclusions}
\citet{Wakelam_2008} argued that it is important to include MN reactions between PAH anions and atomic cations for reliable predictions of elemental abundances in dense molecular clouds. In their work, they followed \citet{Spitzer1941} and used the following expression to calculate the MN rate coefficients in interactions with PAHs with similar sizes as C$_{60}$: 
\begin{equation}
\label{eq.wh}
k^{}_{\mathit{WH}}(T)= \pi R_c^2 v_0 \bigg{(}1+\frac{1}{R_c k_BT}\bigg{)},
\end{equation}
where $k_B$ is the Boltzmann constant and $R_c$ is the critical distance at which the electron is transferred. Here, the mean relative velocity between the collision partners is $v_0= (8k_{B}T/(\pi\mu))^{1/2}$, 
where $T$ is the temperature and $\mu$ the reduced mass. 

In the work by \citet{Wakelam_2008}, final states populated at large electron transfer distances were not considered, as it was assumed that $R_c$ was equal to the geometrical size of the PAH molecule. The expression given by Eq.\ (\ref{eq.wh}) may thus be regarded as a lower limit for the rate coefficient, as it is assumed that R$_c$ is given by geometrical size only. In the present case of C$_{60}^+$ + C$_{60}^-$ collisions, the corresponding minimum (geometrical) value is $R_c^{min}=13.4$\,a$_0$ (twice the fullerene cage radius). If the initial and final states are known, an upper limit can be estimated from the largest curve crossing distance ($R_c^{max}$) for the ion-pair and van der Waals states. Here, $R_c^{max} = 74$\,a$_0$ according to the model potential energy curves shown in Fig.\ \ref{fig:fig3}. The gray area in the bottom part of Fig.\ \ref{fig:fig4} then indicates the possible values of the MN thermal rate coefficient according to Eq.\ (\ref{eq.wh}), with the electron transfer distances between $R_c^{min}$ and $R_c^{max}$. The rate coefficient spans an order of magnitude for a given temperature, which highlights the importance of benchmarking models that take the actual transition probabilities into account in order to provide accurate predictions of MN rates. This is the aim of the present study, where we have shown that the Landau-Zener model, combined with the metal sphere model to describe mutual polarization effects, successfully reproduces the measured final state distribution. In this particular case, we find that the Landau-Zener model rate coefficient is rather close to the $k^{}_{\mathit{WH}}(T)$ curve for the lower limit $R_c^{min}=13.4$\,a$_0$ (see lower part of Fig.\ \ref{fig:fig4}), with the present model predicting rate coefficients that are about 30\% greater than those found in existing models that use this limit \citep{Wakelam_2008}. However, for other collision partners, such as PAHs, the situation may be markedly different, as the reaction cross sections and rates depend sensitively on the Landau-Zener reaction probabilities and, hence, on the quantum nature of the collision partners rather than their geometrical sizes only. In the present case, we find that fitting the rate calculated using the Landau-Zener model to the expression in Eq.\ (\ref{eq.wh}) gives $R_c= 18.5$\,a$_0$, which is in reasonable agreement with the most probable charge transfer distances according to Landau-Zener model calculations (see Fig.\ \ref{fig:fig3}).

The present study of MN in C$_{60}^+$ + C$_{60}^-$ collisions offers a unique opportunity to test the validity of semiclassical models developed for atomic MN and simpler models used for interstellar chemistry on more complex molecular systems. Fullerenes are well suited to benchmark these models, as the inter-molecular interactions may be accurately described by a central potential, and the experimental MN fingerprint (two neutrals in coincidence) does not suffer from contamination from fragmented molecules, which is common for smaller and less stable species~\citep{Poline_2024,bogot2024}. This study thus serves as an important step toward developing semiclassical model frameworks for treating MN involving PAHs and other complex molecules for which orientation-dependent polarization effects are expected to be important and bond-forming reactions may occur. Here, we show that a simple model framework reproduces the measured results remarkably well when the fullerenes are treated as rigid metal spheres and the charge transfer probabilities are calculated using the semiclassical multichannel Landau-Zener method \citep{Bates_1955,Salop_1976}. Further studies will have to be performed to investigate the validity of this approach for other systems or if modifications need to be made, such as describing PAH molecules as planar metal discs instead of as spheres \citep{Linden2016}. The model thermal rate coefficients presented here can be directly applied in astrophysical models of the types used for estimates of the charge balance in interstellar molecular clouds \citep{Wakelam_2008,Bakes1995} and in photo dissociation regions \citep{Millar_2017,Sidhu:2023}. 

The final state populations are important, as they may influence the emission spectra used to identify new molecules and their abundances. For example, the abundance of triplet final states predicted by our calculations (Fig.\ \ref{fig:fig3}) could result in less well studied optically forbidden transitions, as the fullerenes relax to their electronic ground states. In addition, we have shown that the kinetic energy release in MN processes may be substantial due to the strong and long-ranged attraction between the charged collision partners. In this specific case, the typical kinetic energy of each of the neutral C$_{60}$ molecules is $E_{\mathit{KER}}/2=0.65$\,eV, corresponding to a thermal kinetic temperature ($k_BT$) of $\sim 7500$\,K, which may be sufficient to overcome barriers for reactions with other types of molecules where these barriers may be prohibitively high at thermal velocities, such as in cold interstellar molecular clouds.

\begin{acknowledgements}
This work was performed at the Swedish National Infrastructure, DESIREE (Swedish Research Council Contract Nos. 2017-00621 and 2021-00155). H.Z., H.C., H.T.S., and M.G. thank the Swedish Research Council for individual project grants (with contract Nos. 2020-03437, 2023-03833, 2022-02822, and 2020-03104). Furthermore, M.G., R.D.T., H.C., H.T.S., \AA.L. and H.Z.\ acknowledge the project grant "Probing charge- and mass- transfer reactions on the atomic level" (2018.0028) from the Knut and Alice Wallenberg Foundation. This publication is based upon work from the COST Actions CA18212 - Molecular Dynamics in the GAS phase (MD-GAS) and CA21101 - Confined Molecular Systems: from the new generation of materials to the stars (COSY), supported by COST (European Cooperation in Science and Technology).
\end{acknowledgements}

\bibliographystyle{aa} 
\bibliography{references}

\onecolumn
\begin{appendix}

\section{Parameters of cross sections and thermal rate coefficients}

\begin{table*}[h]
    \centering

\caption{Parameters of cross sections and thermal rate coefficients of the processes C$_{60}^+$ + C$_{60}^-$ $\rightarrow$ C$_{60}^*$ + C$_{60}$, C$_{60}^+$ + e$^-$ $\rightarrow$ C$_{60}$, and C$_{60}$ + e$^-$ $\rightarrow$ C$_{60}^-$, according to equations (\ref{eq.x}) and (\ref{eq:rate}), respectively.}
    \label{tab:parameters}
    
    \begin{tabular}
    {lccc}
    \hline \hline
     &  &  &   \\ [-2ex]
Parameter (unit) & C$_{60}^+$ + C$_{60}^-$ & C$_{60}^+$ +e$^-$ & C$_{60}$ +e$^-$ \\ [0.6ex]
\hline 
  &  &  &   \\ [-2ex]
$A$ (cm$^4$s$^{-2}$) & $2.57\times10^{-4}$ & $72.4$ & 0 \\
$B$ (cm$^3$s$^{-1}$) & $1.18\times10^{-11}$ & $9.71\times10^{-7}$ &  $9.71\times10^{-7}$ \\
$C$ (cm$^2$) & $3.44\times10^{-14}$ & $6.51\times10^{-15}$ & $6.51\times10^{-15}$ \\
$D$ (cm s) & $-2.71\times10^{-20}$ & $0$ & $0$ \\
\hline
  &  &  &   \\ [-2ex]
$\alpha_0$ (cm$^3$s$^{-1}$) & $1.18\times10^{-11}$ & $9.71\times10^{-7}$ &  $9.71\times10^{-7}$ \\
$\alpha_1$ (cm$^3$s$^{-1}$K$^{-1/2}$) & $4.57\times10^{-10}$ & $7.00\times10^{-8}$ & $7.00\times10^{-8}$ \\
$\alpha_2$ (cm$^3$s$^{-1}$K$^{1/2}$)& $2.47\times10^{-8}$ & $8.57\times10^{-6}$ & 0 \\
$\alpha_3$ (cm$^3$s$^{-1}$K) & $-5.64\times10^{-12}$ & 0 & 0 \\
    \hline
    \end{tabular}
    
\end{table*}

\section{Vertical excitation energies}
Vertical excitation energies of C$_{60}$ are calculated using TD-DFT with the PBE0 functional~\citep{PBE0} and the \mbox{6-311++G(d,p)} basis set. The energies are provided in Table~\ref{tab:energies} together with branching fractions obtained using the multichannel Landau-Zener model.
\begin{table*}[h]
    \centering
    \caption{Vertical excitation energies $E_{exc}$ in electronvolts for singlet and triplet states of C$_{60}$ and the corresponding branching fractions from the present multichannel Landau-Zener model calculations. The excitation energies are 
    calculated using time-dependent density functional theory (TD-DFT) with the PBE0 functional and the 6-311++G(d,p) basis set. Note that the excitation energies are shifted by +0.15\,eV in the Landau-Zener model calculations, and the branching fractions are calculated at a collision energy of 100\,meV, where the total MN cross section is $5.33\times10^{-13}$ cm$^2$.}
    \begin{tabular}{c c c c c}
    \hline \hline
    \multicolumn{1}{c}{} & \multicolumn{2}{c}{} & \multicolumn{2}{c}{}\\ [-2ex]
\multicolumn{1}{c}{} & \multicolumn{2}{c}{Singlet states} & \multicolumn{2}{c}{Triplet states}\\ [0ex]
 &  & &  & \\ [-2ex]
     State &  $E_{exc}$ (eV) & branching fractions & $E_{exc}$ (eV) & branching fractions\\ [0.4ex]
     \hline 
         1 & 0 & 0.00 & 1.50 & $3.88\times10^{-3}$ \\
         2 & 2.14 & $2.16\times10^{-3}$ & 1.93 &$6.07\times10^{-3}$  \\ 
         3 & 2.32 & $2.70\times10^{-3}$ & 1.97 &$6.24\times10^{-3}$ \\
         4 & 3.3 & $1.33\times10^{-2}$& 2.07 & $6.99\times10^{-3}$\\
         5 & 3.38 & $1.60\times10^{-2}$& 2.44 & $1.13\times10^{-2}$\\
         6 & 3.54 & $2.10\times10^{-2}$& 2.85 & $2.07\times10^{-2}$ \\
         7 & 3.59 & $2.30\times10^{-2}$& 3.04 & $2.90\times10^{-2}$ \\
         8 & 3.62 & $2.55\times10^{-2}$& 3.33 & $5.24\times10^{-2}$ \\
         9 & 3.63 & $2.90\times10^{-2}$& 3.45 & $6.77\times10^{-2}$ \\
         10  & 3.67 & $3.21\times10^{-2}$& 3.48 & $7.59\times10^{-2}$\\
         11 & 3.68 & $3.81\times10^{-2}$& 3.6 & $9.23\times10^{-2}$\\
         12 & 3.75 & $3.88\times10^{-2}$& 3.7 & $9.37\times10^{-2}$\\
         13 & 4.14 & $1.72\times10^{-3}$& 3.71 &$0.107$ \\
         14 & 4.21 & $5.61\times10^{-4}$& 3.79 & $9.52\times10^{-2}$\\
         15 & 4.31 & $7.06\times10^{-5}$& 3.9\ &$5.74\times10^{-2}$ \\ 
         16 & 4.43 & $2.45\times10^{-6}$& 4.01 &$2.43\times10^{-2}$ \\
         17 & 4.54 & $2.67\times10^{-8}$& 4.13 & $5.90\times10^{-3}$\\
         18 & 4.6 & $1.83\times10^{-11}$& 4.34 & $1.02\times10^{-4}$\\
         19 & 4.62 & 0.00& 4.35 &$7.94\times10^{-5}$ \\
         20 & 4.63 & 0.00  & 4.38 & $3.72\times10^{-5}$\\
         21 &4.72 & 0.00 & 4.39 & $2.71\times10^{-5}$\\
         22 & 4.76 & 0.00 & 4.42 & $1.02\times10^{-5}$\\
         23 & & & 4.46 & $2.58\times10^{-6}$ \\
         24 & & & 4.51 & $3.47\times10^{-7}$ \\
         25 & & & 4.59 & $1.90\times10^{-10}$ \\
         26 & & & 4.68 & 0.00 \\
         27 & & & 4.69 & 0.00 \\
         28 & & & 4.71 & 0.00 \\
         29 & & & 4.75 & 0.00 \\
         30 & & & 4.77 & 0.00 \\
         \hline
    \end{tabular}
    \label{tab:energies}
\end{table*}
\end{appendix}

\end{document}